\documentclass[runningheads]{svmult}

\usepackage{makeidx}   
\usepackage{graphicx}  
\usepackage{subeqnar}  
\usepackage{multicol}  
\usepackage{physprbb}  
\makeindex             

\begin{document}
\title*{Evidence for Large Stellar Disks in Elliptical Galaxies.}
\toctitle{Evidence for Large Stellar Disks in Disky and Boxy Elliptical Galaxies.}
%
%
\titlerunning{Stellar Disks in Ellipticals}
%
\author{Andreas Burkert 
\and Thorsten Naab   }
\authorrunning{Andreas Burkert et al.}
%
%
\institute{Max-Planck-Institut f\"ur Astronomie, D-69242 Heidelberg, Germany}

\maketitle              

\begin{abstract}
High-resolution numerical simulations of galaxy mergers
are analysed. The global structure and isophotal shapes of the merger remnants 
are in good agreement with the observations. Whereas equal-mass mergers
lead to anisotropic, boxy ellipticals, unequal-mass mergers result in
disky and isotropic systems.  The line-of-sight velocity distributions 
show small deviations from a Gaussian distribution. In all cases
we find that the retrograde wings are steeper and that the prograde
wings are broader than a Gaussian distribution.
This is in contradiction with the observations which 
show broader retrograde and steeper prograde wings in all ellipticals.
This fundamental difference between observation and theory
can be explained if all ellipticals, even anistropic boxy ones, contain
extended stellar disk components with luminosities of order 10\% to 20\% 
the total luminosity and scale radii of order the effective radii of the spheroids.
\end{abstract}

\section{Introduction}
Elliptical galaxies are believed to form by major mergers of spiral
galaxies \cite{toomre72ab}. Numerous numerical simulations have indeed 
demonstrated that mergers
lead to spheroidal stellar systems with surface brightness profiles 
that are in good agreement with the observations of elliptical galaxies 
(e.g. \cite{barnes88ab,hernquist92ab,hernquist93ab,bekki98ab}).
During the merging epoch the systems are
far from dynamical equilibrium, resulting in
bursts of star formation \cite{mihos96ab,bekki01ab},
the formation of massive star clusters and the infall of gas into
the central regions and the formation and feeding of 
massive central black holes.  When the
merger remnants have settled into dynamical equilibrium,
phase mixing and violent relaxation will erase the information
about the initial conditions. However, as violent relaxation is incomplete,
ellipticals should still show fine structure in their isophotal shape and velocity distribution
which provides insight on their formation.

This conclusion is confirmed by numerical simulations.
Heyl et al. \cite{heyl96ab} showed that the line of sight velocity 
distributions contain information about the initial disk orientations. 
Naab et al. \cite{naab99ab} and Bendo \& Barnes \cite{bendo00ab} demonstrated 
that the isophotal shapes of the remnants 
are determined primarily by the initial mass ratio of the merging galaxies. 

In this paper we focus on signatures for gas infall and star formation during 
the formation of ellipticals. We demonstrate that all ellipticals must 
contain a second disk-like substructure that most likely formed by gas 
infall with subsequent star formation from gas-rich progenitors.

\section{The Line-of-Sight Velocity Distribution of Merger Models}

\begin{figure}[t]
\begin{center}
\includegraphics[width=1.\textwidth]{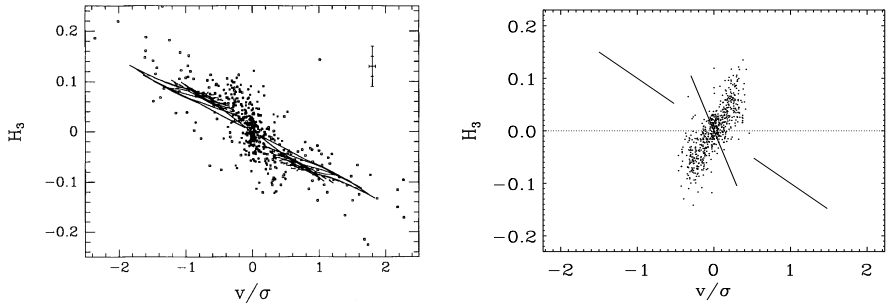}
\end{center}
\caption[]{{\bf Left panel:} Observed local correlation between $H_3$ and $v/\sigma$ 
\cite{bender94ab}. Typical error
bars are plotted in the upper right corner. 
The lines show model predictions for two-integral models 
\cite{dehnen93ab,dehnen94ab}. 
{\bf Right panel:} Correlation beween $H_3$ and $v/\sigma$ for a typical
merger remnant.}
\end{figure}

We have performed very high-resolution N-body simulations with more than $10^5$ particles of 
spiral-spiral mergers with different
mass ratios and orbital parameters. The spirals are constructed in dynamical 
equilibrium using the method described by \cite{hernquist93ab}. Each galaxy 
consists of an exponential disk, a spherical, non-rotating bulge and
a dark halo component. An analysis of the isophotal shapes and the global 
kinematical properties of the merger remnants is presented in \cite{naab99ab}.
We find a strong dependence on the initial mass ratio of the merging components.
Whereas mergers with mass ratios $m_1:m_2 \leq 2:1$ lead to boxy and anisotropic systems,
unequal mass mergers with mass ratios $m_1:m_2 > 2:1$ form disky and isotropic ellipticals.

Another interesting property is the line-of-sight velocity distribution
and its deviations from a Gaussian. Following the procedure discussed by 
\cite{bender94ab} 
we have placed slits with thickness 0.2 the effective radius 
along the apparent long axi of our projected merger remnants. 
The slit is subdivided into grid cells with length 0.15 the effective radius.
All particles within a cell are binned in line-of-sight velocity.
The resulting profile is parametrized using Gauss-Hermite functions 
\cite{bender94ab,bendo00ab}
and the amplitude $H_3$ of the third-order basis function 
is determined by least squares fitting.
In addition, in each grid cell the mean projected radial velocity $v$ and the radial velocity 
dispersion $\sigma$ of all
particles is determined. In agreement with the observations, the deviations 
from Gaussians are small, of order a few percent

\begin{figure}[t]
\begin{center}
\includegraphics[width=0.85\textwidth]{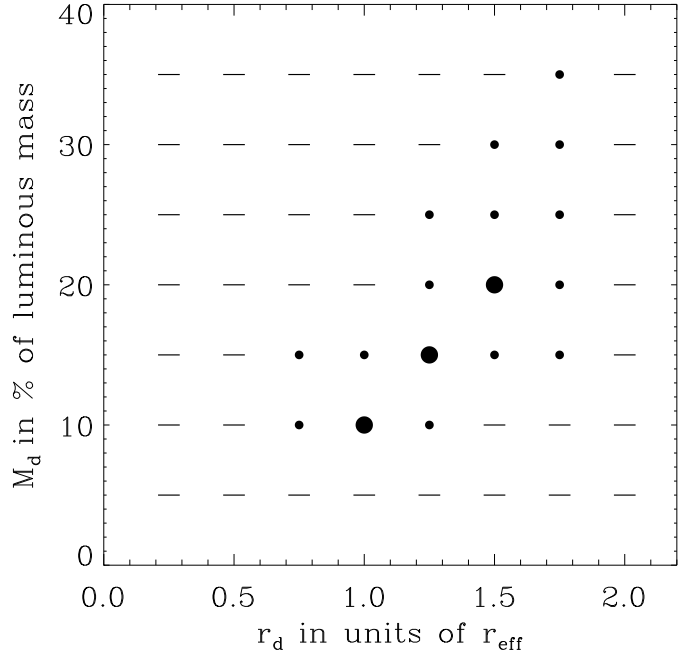}
\end{center}
\caption[]{Values of disk masses $M_d$ versus disk scale lengths $r_d$
that have been added to a merger remnant.
Big dots represent solutions that provide an excellent fit to the observed
$H_3$ versus $v/\sigma$ relationship. Small dots show
values that provide a reasonable fit. Minus signs show combinations that fail.}

\end{figure}

\section{Comparison with Observations}

Figure 1 (left panel) shows the local correlation between 
$H_3$ and $v/\sigma$ as measured 
by \cite{bender94ab} for their observed sample of galaxies. In all cases $H_3$ 
and $v/\sigma$ have opposite signs, that is
the prograde wings of the line-of-sight velocity distributions are always steeper
than the retrograde wings. This  result holds not only for ellipticals in low-density 
environments studied by \cite{bender94ab} but also for ellipticals in the 
Coma cluster \cite{mehlert00ab}. The right panel of Figure 1 shows the 
result of a representative merger simulation,
an unequal mass merger which leads to a disky, isotropic elliptical.
The correlation between $v/\sigma$ and $H_3$ 
is opposite to the observed one.  The profiles have broad prograde 
wings and narrow retrograde wings. The same signature is found in all remnants,
independent of whether they are disky or boxy,  isotropic or anisotropic.
The only exception are equal mass mergers of counter-rotating disks which lead to
very anisotropic ellipticals with no signature of rotation.

\section{Disk-like Subcomponents in Elliptical Galaxies}

One possible solution is the existence of a second disk component.
In order to test this assumption we have placed stellar disks in the equatorial plane 
of the merger remnants. The disks rotate 
in centrifugal equilibrium with the total gravitational potential. Adopting an
exponential disk surface brightness profile, the only free parameters are the
ratio of disk mass to spheroid mass and the scale length of the disk, 
normalized to the scale radius of the spheroid.  The results are 
summarized in Figure 2. Disks with 
small masses or radii do not change the line-of-sight velocity distribution of the stellar 
component, especially in the outer regions where $v/\sigma$ is large.
Disks with mass ratios and radii in the region shown by the 
filled dots in Figure 2, on the other hand,
lead to a significant change in the line profiles. The prograde wings become steeper 
than the retrograte ones also at an effective radius and beyond,
in very good agreement with the observations. If the disks become 
too massive, the surface brightness profiles of the spheroids change from
de Vaucouleurs profiles to exponential profiles which is again not in agreement
with the observations.
We therefore can conclude that disks with 10\% to 20\% the luminosity of the spheroids and 
scale lengths of order the effective radii of the spheroids
can explain the observed correlation between $H_3$ and $v/\sigma$ in ellipticals.

The origin of extended disk components in ellipticals is not understood 
up to now. Bekki \cite{bekki00ab} discussed the formation of nuclear disks 
in mergers which 
however have scale lengths that are small compared to the length scales of the spheroids.
We recently started a new set of merger simulations of very gas-rich spiral galaxies.
For unequal mass mergers, which lead to disky ellipticals, the 
gas settles indeed into extended disks if star-formation is suppressed 
during the early merging phase.  However, in the case of equal-mass mergers, which produce
boxy ellipticals, tidal torques lead to efficient angular momentum loss in the gaseous 
component, resulting  in gas infall into the center and the formation  and growth
of central massive black holes. No extended gaseous disks are formed in this case.
The origin of large stellar disks in boxy ellipticals is therefore still unclear.

%


\begin{thebibliography}{8.}
\addcontentsline{toc}{section}{References}

\bibitem{barnes88ab} J.E. Barnes: ApJ {\bf 331}, 699 (1988)
\bibitem{bekki98ab} K. Bekki: ApJ {\bf 502}, L133 (1998)
\bibitem{bekki00ab} K. Bekki: ApJ {\bf 545}, 753 (2000)
\bibitem{bekki01ab} K. Bekki: ApJ {\bf 546}, 189 (2001)
\bibitem{bender94ab} R. Bender, R.P. Saglia, O.E. Gerhard: 
  MNRAS {\bf 269}, 785 (1994)
\bibitem{bendo00ab} G.J. Bendo, J.E. Barnes: MNRAS {\bf 316}, 315 (2000)
\bibitem{dehnen93ab} W. Dehnen, O.E. Gerhard: MNRAS {\bf 261}, 311 (1993)
\bibitem{dehnen94ab} W. Dehnen, O.E. Gerhard: MNRAS {\bf 268}, 1019 (1994)
\bibitem{hernquist92ab} L. Hernquist: ApJ {\bf 400}, 460 (1992)
\bibitem{hernquist93ab} L. Hernquist: ApJ {\bf 409}, 548 (1993)
\bibitem{heyl96ab} J.S. Heyl, L. Hernquist, D.N. Spergel: 
 ApJ {\bf 463}, 69 (1996)
\bibitem{mehlert00ab} D. Mehlert, R.P. Saglia, R. Bender, G. Wegner: 
 A\&AS {\bf 141}, 449 (2000)
\bibitem{mihos96ab} J.C.~Mihos, L.~Hernquist: ApJ {\bf 464}, 641 (1996)
\bibitem{naab99ab} T. Naab, A. Burkert, L. Hernquist: 
 ApJ {\bf 523}, L133 (1999)
\bibitem{toomre72ab} A. Toomre, J. Toomre: ApJ {\bf 178}, 623 (1972)
\end{thebibliography}
\end{document}